\documentclass[preprint,preprintnumbers,amsmath,amssymb,nofootinbib]{revtex4}
\usepackage{graphicx}
\usepackage{bm}
\newcommand\nn{\nonumber \\}
\newcommand\vk{{\vec{k}}}

\newcommand\vx{{\vec{x}}}

\newcommand\bq{{\bf q}}
\newcommand\br{{\bf r}}
\newcommand\bnab{{\bf \nabla}}

\def\sla{\slash{\!\!\!}}

\newcommand{\lk}{\left(}
\newcommand{\rk}{\right)}
\newcommand{\ltk}{\left\{}
\newcommand{\rtk}{\right\}}
\newcommand{\ldk}{\left[}

\newcommand{\rdk}{\right]}
\newcommand\beq{ \begin{eqnarray} }
\newcommand\eeq{ \end{eqnarray} }

\begin{document}


\title{Landau-Peierls instability in a Fulde-Ferrell type inhomogeneous chiral condensed phase}
\author{Tong-Gyu Lee}%
\email{tonggyu.lee@yukawa.kyoto-u.ac.jp}%
\affiliation{Yukawa Institute for Theoretical Physics, Kyoto University, Kyoto 606-8502, Japan}
\author{Eiji Nakano}
\email{e.nakano@s.kochi-u.ac.jp}
\author{Yasuhiko Tsue}
\email{tsue@kochi-u.ac.jp}
\affiliation{Department of Physics, Kochi University, Kochi 780-8520, Japan}
\author{Toshitaka Tatsumi}%
\email{tatsumi@ruby.scphys.kyoto-u.ac.jp}%
\affiliation{Department of Physics, Kyoto University, Kyoto 606-8502, Japan}
\author{Bengt Friman}
\email{b.friman@gsi.de}
\affiliation{GSI, Helmholzzentrum f\"ur Schwerionenforschung, Planckstrasse 1, D-64291 Darmstadt, Germany}

\date{\today}

\begin{abstract}
We investigate the stability of an inhomogeneous chiral condensed phase 
against low energy fluctuations about a spatially modulated order parameter. 
This phase corresponds to 
the so-called dual chiral density wave in the context of quark matter, 
where the chiral condensate is spatially modulated with a finite wavevector in a single direction.
From the symmetry viewpoint, 
the phase realizes a locking of flavor and translational symmetries. 
Starting with a Landau-Ginzburg-Wilson effective Lagrangian, 
we find that the associated Nambu-Goldstone modes, whose dispersion relations are spatially anisotropic and soft in the direction normal to the wavevector of the modulation, 
wash out the long-range order at finite temperatures, but support algebraically decaying long-range correlations. 
This implies that the phase can exhibit a quasi-one-dimensional order 
as in liquid crystals. 
\end{abstract}


\maketitle

\section{Introduction}
\label{sec:1}
Unveiling the phase diagram of quantum chromodynamics (QCD) is among the most fundamental issues in quark-hadron physics. 
So far, considerable theoretical and experimental efforts have been devoted to exploring the QCD phase diagram (for reviews see Ref.~\cite{Fukushima:2010bq});  
the properties of the high temperature regime are studied 
experimentally in ultra-relativistic heavy-ion collisions, 
and in ab-initio lattice QCD simulations. 
The latter are subject to technical difficulties at non-zero net baryon-number density, the so-called sign problem.
In the near future, data at lower beam energies, relevant for the exploration of the phase diagram at non-vanishing
baryon density, will be forthcoming. In order to exploit this opportunity in an optimal way, it is necessary 
to find appropriate observables for deciphering the properties of dense and moderately hot matter in such collisions~\cite{Friman:2011zz}.

In recent theoretical studies of QCD at finite temperature and density, 
various inhomogeneous chiral condensed phases have been proposed 
(for a recent review see Ref.~\cite{Buballa:2014tba}).
These studies suggest that the conventional QCD phase diagram should be redrawn.
Indeed, it is possible that the phase structure at high net baryon densities and moderate temperatures is modified considerably 
by the presence of inhomogeneous phases. Thus, the region of the chiral transition may be extended and the order of the phase transition may change.  
These features are gleaned primarily from mean-field calculations 
in the Nambu-Jona-Lasinio (NJL) and quark-meson (QM) models~\cite{Nakano:2004cd,Nickel:2009wj}, 
and the Dyson-Schwinger approach to dense QCD~\cite{Muller:2013tya}. 
It also interesting to note that within the Gor'kov approach to chiral effective models~\cite{Nickel:2009ke,Carignano:2014jla}
it is found that the QCD critical endpoint is a Lifshitz point, 
where the normal, homogeneous, and inhomogeneous chiral condensed phases meet.
In the large $N_c$ approach to dense QCD, 
early studies suggested the emergence of the so-called chiral density wave~\cite{Deryagin:1992rw,Shuster:1999tn}, 
while in the context of quarkyonic matter~\cite{McLerran:2007qj} 
another inhomogeneous phase, the so-called quarkyonic chiral spiral, was discussed~\cite{Kojo:2009ha}. 

The inhomogeneous chiral condensed phases mentioned above  
correspond to a one-dimensional modulation embedded in three spatial dimensions. Some 
of these structures are based on extrapolations from 
analytic solutions obtained for purely (1+1)-dimensional systems~\cite{Schon:2000qy}. 
Possible extensions to higher-dimensional modulations have been studied,
with the result that the one-dimensional modulation tends to be favored close to the Lifshitz point~\cite{Abuki:2011pf} and/or at zero temperature~\cite{Carignano:2012sx}.

Let us start by classifying the modulations for inhomogeneous phases, 
according to the convention employed in condensed matter physics. 
There are basically two types of one-dimensional modulations: 
one is of the Fulde-Ferrell (FF) type~\cite{Fulde:1964zz}, 
characterized by modulations of the phase of a complex order parameter with constant amplitude, 
while the other is of the Larkin-Ovchinnikov (LO) type~\cite{larkin:1964zz}, 
where by contrast only the amplitude is modulated. 
The FF type includes the dual chiral density wave (DCDW)~\cite{Nakano:2004cd} 
and the quarkyonic chiral spiral~\cite{Kojo:2009ha}. 
On the other hand, 
the chiral density wave (a plane wave)~\cite{Deryagin:1992rw,Shuster:1999tn}
and periodic domain walls~\cite{Nickel:2009wj} are of the LO type. 
In the present paper we focus on the DCDW, which is of FF type.  
The DCDW is characterized by modulated scalar and pseudoscalar condensates 
with a constant amplitude $\Delta$ and a wavenumber $q$~\cite{Nakano:2004cd}, 
\beq
\langle \bar{\psi}\psi \rangle =\Delta \cos{qz}, \quad 
\langle \bar{\psi}i\gamma_5\tau_3\psi \rangle =\Delta \sin{qz},  
\eeq
where $\psi$ is the quark field for two flavors, 
and $\tau_3$ a Pauli matrix diagonal in the isospin space. 
This configuration is akin to 
$\sigma\pi^0$ condensation, obtained in neutron matter within the sigma model~\cite{Dautry:1979bk}, 
and is thus expected to smoothly connect between nuclear and quark matter.

Most studies of inhomogeneous chiral condensed phases so far are based on mean-field calculations. 
Thus, effects of thermal and quantum fluctuations have yet to be studied. 
In the context of pion condensation, 
the stability of modulated condensates against thermal fluctuations has been studied on the basis of Landau-Peierls arguments~\cite{landau1969statical}.   
It was found that in such systems there is no true long-range order with a non-vanishing order parameter~\cite{Baym:1982ca}. Instead, such systems can develop a quasi-ordered one-dimensional condensate, with correlation functions that decay algebraically in space.%
\footnote{A similar state is found, e.g., in smectic liquid crystals~\cite{AlsNielsen:1980zz,de1993physics}. See also Refs.~\cite{Shimahara1998,Radzihovsky2009} for a corresponding discussion on FFLO superconductors/superfluids.}
In this paper, we investigate the stability of the DCDW phase 
against low-energy fluctuations of Nambu-Goldstone (NG) modes associated with the spontaneous symmetry breaking, along the lines of Ref.~\cite{Baym:1982ca}.

The paper is organized as follows.
In the next section, 
we construct 
a (3+1)-dimensional Landau-Ginzburg-Wilson effective Lagrangian 
for general order parameters of the chiral condensate, 
which are allowed to be spacetime dependent, 
and then apply the formalism to the DCDW phase. 
In Sec.~\ref{sec:3}, 
we discuss the symmetry breaking pattern and the corresponding NG modes in the DCDW phase.
We also present the dispersion relations for these low-energy collective excitation modes, by introducing fluctuations such as amplitudons and phonons/phasons (NG modes) on the ground state of the DCDW.
In Sec.~\ref{sec:4}, 
we investigate the impact of low-energy fluctuations on the order parameter by evaluating the long-range correlation functions.
Finally, section~\ref{sec:5} is devoted to summary and outlook.

\section{Landau-Ginzburg-Wilson effective Lagrangian}
\label{sec:2}
We start by introducing a $2\times 2$ matrix field $M$ as the $\lk {1}/{2},{1}/{2}\rk$ representation of the chiral $SU(2)_L \times SU(2)_R$ symmetry. In the following, we shall use the fact that $SU(2)_L \times SU(2)_R$ is isomorphic to the four-dimensional rotation group $O(4)$.
The matrix $M$ can be expressed in terms of the right and left handed quark fields $\psi_{R,L}$~\cite{lee1972chiral},
$M_{ij}= {\psi}_{L,i}\psi_{R,j}^{\dagger}$.
Under the transformations $U_{R,L}=\exp\ldk -i\lk \vec{\alpha}\pm\vec{\beta}\rk \cdot\vec{\tau}/2 \rdk$, where $\vec{\tau}$ is the isospin Pauli matrix  and $\vec{\alpha}$ and $\vec{\beta}$ are two three-dimensional vector parameters,
the matrix $M$ transforms (to leading order in $\vec{\alpha}$ and $\vec{\beta}$) as $M_{ij} \rightarrow M_{ij}+\frac{i}{2}\vec{\alpha}\cdot\ldk\vec{\tau},M\rdk_{ij} +\frac{i}{2}\vec{\beta}\cdot\ltk \vec{\tau},M\rtk_{ij}$.
With the parameterization $M = \sigma + i\vec{\pi}\cdot \vec{\tau}$, one finds the corresponding transformation laws for $\sigma$ and $\vec{\pi}$: $\sigma \rightarrow \sigma - \vec{\beta}\cdot \vec{\pi}$ and
$\vec{\pi} \rightarrow \vec{\pi}-\vec{\alpha}\times \vec{\pi}+\vec{\beta}\,\sigma$.
Thus, the rotation with $\vec{\alpha}$ corresponds to the vector (isospin) rotation while that with $\vec{\beta}$ to an axial vector (chiral or axial isospin) rotation, respectively. We can then introduce a four-component composite field
$\phi^T = \lk \sigma, \vec{\pi} \rk$, which transforms as a four-dimensional vector under $O(4)$ rotations.

Now, we construct an effective Lagrangian density ${\mathcal L}$ with $O(4)$ symmetry in terms of $\phi$ and its derivatives,%
\footnote{For non-zero isospin charges, 
	a term of the form $\frac{1}{2} \epsilon_{ijkl} \mu_{ij} Q_{ij}$ is added to the Lagrangian. Here $\epsilon_{ijkl}$ ($\epsilon_{1234}=1$) is the antisymmetric tensor, and $\mu_{ij}=-\mu_{ji}$ the chemical potential for the isospin density 
	$Q_{ij}=i(\phi_i\partial_0 \phi_j-\phi_j\partial_0 \phi_i)$.}
\beq
{\mathcal L}&=& 
c_2\partial_0\phi\cdot\partial_0\phi -{\mathcal V}, \label{Lag} \\ 
{\mathcal V}&=& a_2\phi\cdot \phi+a_{4,1}\lk \phi\cdot\phi\rk^2
+a_{4,2}\bnab \phi\cdot\bnab \phi
+a_{6,1}\bnab^2 \phi\cdot\bnab^2 \phi\nn
&&+a_{6,2}\lk \bnab \phi\cdot\bnab \phi\rk\lk \phi\cdot\phi\rk
+a_{6,3}\lk\phi\cdot\phi\rk^3
+a_{6,4}\lk \phi\cdot \bnab \phi\rk^2, 
\eeq
Note that since the Lagrangian describes the excitations of a medium, we do not assume Lorentz invariance.
The potential term ${\mathcal V}$ is expanded up to sixth order in powers of the field and fourth order in its derivatives, 
as required for stability of the inhomogeneous phase at the mean-field level.%

Hereafter we set $c_2=1$ 
for simplicity. 
The expansion coefficients $a_{i,j}$ in ${\mathcal V}$ can be evaluated within effective chiral models, like the NJL~\cite{Nickel:2009ke} and QM~\cite{Nickel:2009wj,Carignano:2014jla} models. In the former, one finds the following relations among them $a_{4,1} = a_{4,2}$ and $\lk a_{6,1}, a_{6,2}, a_{6,4} \rk = \lk {1}/{2}, 3, 2 \rk a_{6,3}$. 

\section{Low energy effective modes in DCDW phase}
\label{sec:3}
We consider an inhomogeneous time independent chiral condensate of the DCDW type, 
\beq
\phi_0^T = \Delta \lk \cos{qz}, 0,0, \sin{qz}\rk, 
\eeq
where $\Delta$ is a constant amplitude corresponding to 
$\langle \bar{\psi}e^{i\gamma_5 \tau_3 \bq\cdot\br} \psi \rangle$
and $q$ is the wavenumber of modulation in the $z$ direction.  
The values of $\Delta$ and $q$ are determined by minimizing the potential term of the Lagrangian. 
For the condensate $\phi_0$, the potential term  reads 
\beq
\mathcal{V}(\phi_0)=
a_2 \Delta^2+a_{4,1} \Delta^4
+a_{4,2} q^2\Delta^2+a_{6,1} q^4\Delta^2+a_{6,2} q^2\Delta^4
+a_{6,3}\Delta^6.
\eeq
Stability of the inhomogeneous phase is guaranteed by
\beq
a_{6,1}>0 \ ({\rm or}\ a_{6,3}>0), 
\quad {\rm and} \quad a_{6,1}a_{6,3}-a_{6,2}^2/4>0. 
\eeq
The stationary conditions for $q$ and $\Delta$,
$\frac{\partial {\mathcal V}}{\partial \Delta}=\frac{\partial {\mathcal V}}{\partial q}=0$, yield 
\beq
&&2q\Delta^2 \lk a_{4,2}+2a_{6,1}q^2+a_{6,2}\Delta^2 \rk =0, \\  
&&2\Delta \ldk a_2+a_{4,2}q^2+a_{6,1}q^4+2 \lk a_{4,1}+a_{6,2}q^2 \rk \Delta^2
+3a_{6,3}\Delta^4 \rdk =0. 
\eeq
They admit three types of solutions: 
\begin{enumerate}
\item[(a)] Normal phase: $q=\Delta=0$, 
\item[(b)] Homogeneous chiral condensed phase: $q=0$, $\Delta\neq0$,
\item[(c)] Inhomogeneous chiral condensed phase: 
$q^2=-\lk a_{4,2}+a_{6,2} \Delta^2\rk/2a_{6,1}$, $\Delta\neq0$.
\end{enumerate}
For a given set of coefficients, the phase with the lowest energy, is realized on the classical level. 
The coefficients implicitly depend on the medium, and are thus functions of thermodynamic 
variables, like temperature and chemical potentials.

\subsection{Symmetry breaking and Nambu-Goldstone modes in DCDW phase}\label{subsec:SSBNG}

In the DCDW phase with non-vanishing $\Delta$ and $q$, the $SU(2)_L \times SU(2)_R$ chiral symmetry, as well as the translational invariance in the $z$ direction and the symmetry under rotations about the $x$ and $y$ axes are spontaneously broken. 
To see the symmetry breaking pattern explicitly, we first perform infinitesimal transformations corresponding to a spatial translation in the $z$ direction with a displacement parameter $s$ and a chiral 
rotation through the angles $\vec{\alpha}$ and $\vec{\beta}$:  
\beq
\phi_0 \rightarrow  \phi_0 + 
\Delta \lk 
\begin{array}{c}
-\lk sq+\beta_3 \rk \sin{qz}\\
\beta_1\cos{qz}-\alpha_2\sin{qz} \\ 
\beta_2\cos{qz}+\alpha_1\sin{qz} \\
\lk sq+\beta_3\rk \cos{qz}
\end{array}
\rk. 
\label{SSB1}
\eeq
The form of the first and the fourth components implies that $\phi_0$ is invariant under
a simultaneous spatial translation and axial isospin rotation about the $z$ axis through the angle $\beta_3$, if $qs+\beta_3=0$. Thus, in the DCDW phase a locking of axial isospin rotations with translations is realized. 
In terms of symmetry generators, there are two unique orthogonal linear combinations of $s$ and $\beta_{3}$; one corresponding to a broken generator, the other to an unbroken one.
Consequently, the corresponding NG mode is generated by a transformation with $qs+\beta_3\neq 0$, i.e., 
by one whose generator is broken in the DCDW phase. In the following, we use $\beta_3=\beta_3(t, \vx)$ 
and $s=0$ to generate the NG mode associated with the broken generator. Similar arguments for the 
spontaneous breakdown of internal and spacetime symmetries are given in Refs.~\cite{Kobayashi:2014xua,Hidaka:2014fra}.

The rotations through $\beta_1$ and $\alpha_2$ generate variations 
in the second component in (\ref{SSB1}).
However, the corresponding NG modes are linearly dependent in the sense discussed in Ref.~\cite{Low:2001bw}. 
In case of a vanishing wavenumber $q=0$, only $\beta_1$ is relevant. Thus, we generate the corresponding NG mode
using $\beta_1=\beta_1(t,\vx)$ and $\alpha_2=0$. Analogous arguments can be applied the third component in (\ref{SSB1}), thus eliminating $\alpha_1$ in favour of $\beta_2$.

Spatial rotations about the $x$-axis by an angle $\theta_x$ yields a transformation, 
which is nonuniform in space: $z \rightarrow z \cos\theta_x + y\sin\theta_x$. 
Similarly, the rotations about the $y$-axis  by $\theta_y$ yields the analogous transformation. 
However, the corresponding NG modes and those generated by translations are also linearly dependent~\cite{Low:2001bw}. 

We conclude that, although there are eight broken generators for internal and space time symmetries in the DCDW phase, only three independent NG modes 
remain. These can be chosen as the axial isospin rotations generated by $\vec{\beta}=\vec{\beta}(t,\vx)$.

\subsection{Low energy collective excitations}\label{subsec:LECE}
We now consider a general fluctuation in the DCDW phase:
\beq\label{DCDW-gen-fluct}
\phi&=&(\Delta +\delta)
\lk 
\begin{array}{l}
\cos{\lk qz+\beta_3\rk}\cos{\beta_2}\cos{\beta_1} \\
\cos{\lk qz+\beta_3\rk}\cos{\beta_2}\sin{\beta_1} \\
\cos{\lk qz+\beta_3\rk}\sin{\beta_2} \\
\sin{\lk qz+\beta_3\rk}
\end{array}
\rk 
=(\Delta +\delta) U(\beta_i)
\lk 
\begin{array}{c}
\cos{\lk qz\rk}\\
0\\
0\\
\sin{\lk qz\rk}
\end{array}
\rk.
\eeq
Here $\delta$ is the amplitude fluctuation,
the parameters $\vec{\beta}=\ltk \beta_1,\beta_2,\beta_3\rtk$ 
specifies a rotation in the four-dimensional space spanned by the $\sigma$ and $\vec{\pi}$ fields. Finally, $U(\beta_{i=1,2,3}):=e^{i\beta_1 L_1}e^{i\beta_2 L_2}e^{i\beta_3 L_3}$, where $L_{1,2,3}$ are the O(4) (axial isospin) generators~\cite{lee1972chiral}.
This parametrization clearly shows that the displacement in the $z$ direction is equivalent to a chiral rotation through $\beta_{3}$. To leading order in the fluctuations, (\ref{DCDW-gen-fluct}) yields 
\beq
\phi = \lk 1+\delta \rk \phi_0 + \Delta
\lk 
\begin{array}{c}
-\beta_3\sin{qz}\\
\beta_1\cos{qz}\\
\beta_2\cos{qz}\\
\beta_3\cos{qz}\\
\end{array}
\rk +{\mathcal{O}}\lk \beta_i^2, \delta\beta_i, \delta^2\rk,
\eeq
which exhibits the fluctuation of the amplitude in addition to the fluctuations corresponding to the NG modes.
In the following we consider local fluctuations, promoting the parameters $\delta$ and $\vec{\beta}$ to fields $\delta(x)$ and $\vec{\beta}(x)$, where we use the compact notation $x\equiv \{t, \vx \}$. 

Plugging the above parametrization into the Lagrangian, we can systematically derive a low energy effective theory by expanding 
in powers of the fluctuation fields $\delta$ and $\vec{\beta}$.
Up to the second order in the fields,
the Lagrangian $L=\int {\rm d}^3x{\mathcal L}$ reads 
\beq
{\mathcal L}&=&
\lk \partial_0 \delta\rk^2+\Delta^2(\partial_0 \vec{\beta}_U)^2 + \Delta^2(\partial_0 \beta_3)^2 - \lk {\mathcal V}_\delta+{\mathcal V}_{\delta \beta}+{\mathcal V}_\beta\rk,
\eeq
where
\beq
{\mathcal V}_\delta
&=& M^2 \delta^2
+a_{6,4}\Delta^2(\nabla\delta)^2 +4a_{6,1}q^2(\nabla_z\delta)^2 +a_{6,1}(\nabla^2\delta)^2, 
\\
{\mathcal V}_{\delta \beta} 
&=& 4q\Delta \ldk  a_{6,2}\Delta^2\delta -2a_{6,1}\nabla^2\delta\rdk  \nabla_z\beta_3, 
\\
{\mathcal V}_\beta
&=&
a_{6,1}\Delta^2\lk \nabla^2\vec{\beta}_U +q^2\vec{\beta}_U\rk^2 +a_{6,1}\Delta^2\ldk \lk \nabla^2\beta_3\rk^2 +4q^2\lk \nabla_z\beta_3\rk^2 \rdk, 
\eeq
with the mass term $M^2=4(a_{4,1}+a_{6,2}q^2)\Delta^2 +12a_{6,3}\Delta^4$ and the transverse field $\vec{\beta}_U\equiv \vec{\beta}_T\cos{qz}$ where $\vec{\beta}_T = \ltk \beta_1,\beta_2\rtk$. 
In the above equations the stationary condition $a_{4,2}+a_{6,2}\Delta^2+2q^2a_{6,1}=0$ has been used.
For details of the derivation, see Appendix~\ref{app:A}.

In order to investigate the thermodynamics of the system, we now move to Euclidean space: 
$t \rightarrow -i\tau$ with the period $0\le \tau \le \beta$ 
where $\beta=1/T$ is the inverse temperature. 
For Gaussian fluctuations, 
we obtain the Euclidean action in Fourier space 
(for details we refer the reader to Appendix~\ref{app:B}),
\beq
-S_{E}
&=& 
\int_0^\beta {\rm d}\tau \int {\rm d}^3 x {\mathcal L}_E
=
\sum\!\!\!\!\!\!\!\int{\rm d}k
\lk 
\begin{array}{c}
\delta^*(k)\\
\Delta \beta_3^*(k)
\end{array}
\rk^T
\lk 
\begin{array}{cc}
S_{\delta 0}^{-1}(k) & -g(k) \\
g(k) & S_{0}^{-1}(k)
\end{array}
\rk
\lk 
\begin{array}{c}
\delta(k)\\
\Delta \beta_3(k)
\end{array}
\rk 
\nn
&+&
\frac{1}{4} 
\sum\!\!\!\!\!\!\!\int{\rm d}k
\lk 
\begin{array}{c}
\Delta\vec{\beta}^*_T(k)\\
\Delta\vec{\beta}^*_T(k+2q\hat{z})
\end{array}
\rk^T
\lk
\begin{array}{cc}
S_0^{-1}(k) & G(k) \\
G(k) & S_0^{-1}(k+2q\hat{z})
\end{array}
\rk
\lk 
\begin{array}{c}
\Delta\vec{\beta}_T(k)\\
\Delta\vec{\beta}_T(k+2q\hat{z})
\end{array}
\rk, \nn 
\eeq
where we have used the shorthand notation: 
$\Sigma\!\!\!\!\!\!\!~\int{\rm d}k \equiv T\sum_n \int \frac{{\rm d}^3 k}{(2\pi)^3}$,
and $k=(\omega_n, \vec{k})$ with the Matsubara frequency $\omega_n=2\pi n T (\equiv i\omega)$.
The inverse propagators in the above matrix notation are given by 
\beq
S_{\delta0}^{-1}(k)
&=&\omega^2-\ldk 
M^2 + {a_{6,4}\Delta^2\vec{k}^2}
+4a_{6,1}q^2(k_z)^2+a_{6,1}(\vec{k}^2)^2\rdk, \nn
g(k)
&=&
2iq\ldk  a_{6,2}\Delta^2
+2a_{6,1}\vec{k}^2\rdk k_z, \nn
S_0^{-1}(k)
&=&
\omega^2 
-a_{6,1}\ldk 4q^2 k_z^2 +(\vec{k}^2)^2 \rdk, 
\nn
\mbox{and} \quad 
G(k)
&=&
\omega^2 
-a_{6,1}\lk \vec{k}^2+2qk_z \rk^2. \label{disp}
\eeq
We note that for a non-vanishing wavenumber $q$, the $\delta$ and $\beta_3$ fluctuations mix. Moreover, transverse fluctuations $\vec{\beta}_T$ with different momenta $k$ and $k+2q\hat{z}$ mix, owing to the scattering of fluctuations off the background modulation. 

The determinant of the first matrix, $S_{\delta0}^{-1}(k)S_{0}^{-1}(k)+g^2(k)=0$, yields the dispersion relations of the normal modes involving $\delta$ and $\beta_3$,  
\beq
\omega_+^2&\simeq&M^2+a_{6,1}\ldk u^2_{z+} k_z^2+\lk\vec{k}^2\rk^2\rdk +a_{6,4}\Delta^2\vec{k}^2+A \vec{k}^2k_z^2 +Bk_z^4, \\
\omega_-^2&\simeq&a_{6,1}\ldk u^2_{z-} k_z^2+\lk\vec{k}^2\rk^2\rdk -A \vec{k}^2k_z^2 -Bk_z^4, \label{disp-}
\eeq
where $u_{z\pm}^2= 4q^2\lk 1\pm\frac{a_{6.2}^2\Delta^4}{a_{6.1}M^2}\rk$,
$A\equiv 4q^2\Delta^2 a_{6.2}\lk 4 M^2 a_{6.1}-\Delta^4a_{6.2}a_{6.4}\rk/M^4$, and 
$B\equiv -\lk 2q\Delta^2 a_{6.2}\rk^4/M^6$.
Note that in the massless mode $\omega_-$, 
the dependence on the transverse momentum is subleading, $\mathcal{O}(k^4)$.%
\footnote{The sign of $u^2_{z-}$ in $\omega_-$ is always positive if $a_{4.1}>0$, which generally holds if the transition is second order.}
Consequently, the transverse fluctuations are softer that the longitudinal ones.

Similarly, equating the determinant of the second matrix to zero, 
$S_{0}^{-1}(k)S_{0}^{-1}(k)-G^2(k)=0$,
we obtain the dispersion relation for $\vec{\beta}_T$, 
\beq
\omega_k^2=a_{6,1}\ldk 4q^2 k_z^2+\lk \vec{k}^2\rk^2\rdk 
-a_{6,1}\frac{2k_z^2\lk \vec{k}^2\rk^2}{4q^2+6qk_z+2k_z^2+\vec{k}^2}. 
\label{disp1}
\eeq
The second term with the negative sign is a higher order correction of ${\mathcal O}(k^6)$, stemming from interactions with the background modulation. This term is irrelevant for the effects of low energy fluctuations and is therefore 
dropped in the following discussion.

\section{Impacts of low energy fluctuations}
\label{sec:4}
At low temperatures, the low energy fluctuations about the calssical DCDW state dominate.
We evaluate the contribution of Gaussian fluctuations to the partition function
\beq
Z = \int \left[{\mathcal D}\delta\right] \left[{\mathcal D}\Delta\vec{\beta}\right] e^{-S_E}.
\eeq
Higher-order derivative corrections are dropped, with the assumption that fluctuations 
at energies above some cutoff $\Lambda$ have been integrated out in the effective Lagrangian (\ref{Lag}),
which then involves only the low-energy fluctuations, $\delta$ and $\vec{\beta}$, explicitly.

We first explore the impact of low energy fluctuations on the order parameter,  
\beq
\langle (\Delta+\delta) U(\beta_i)\,\phi_0 \rangle 
&=& 
\Delta \langle U(\beta_i)\,\phi_0 \rangle +\langle \delta\, U(\beta_i)\,\phi_0 \rangle, 
\label{op1}
\eeq
where we use the compact notation 
$\langle \cdots \rangle \equiv 
\int \left[{\mathcal D}\delta\right] \left[{\mathcal D}\Delta\vec{\beta}\right] \cdots e^{-S_E}/Z$. 
In the Gaussian approximation, the two contributions to the expectation value reduce to  
\beq
\langle U(\beta_i)\,\phi_0 \rangle 
&\simeq&
\lk 
\begin{array}{l}
\cos(qz)e^{-\sum_i\langle \beta_i^{2} \rangle/2}  \\
0 \\
0\\
\sin(qz)e^{-\langle \beta_3^{2} \rangle/2} 
\end{array}
\rk
\\
{\rm and}&&\nonumber
\\
\langle \delta\, U(\beta_i)\,\phi_0 \rangle 
&\simeq &
\lk 
\begin{array}{l}
-\sin(qz)\langle \delta \beta_3\rangle e^{-\sum_i\langle \beta_i^{2} \rangle/2}  \\
0 \\
0\\
\cos(qz)\langle \delta \beta_3\rangle e^{-\langle \beta_3^{2} \rangle/2} 
\end{array}
\rk . \qquad
\eeq
Here the second order fluctuations are given by 
\beq
&&\langle \delta(x) \beta_3(x)\rangle
\simeq 0, 
\\
&&
\Delta^2\langle \beta_{1,2}^2(x)\rangle
\simeq \frac{1}{2}\int \frac{{\rm d}^3k}{(2\pi)^3}\frac{T}{\omega_k^2}, \quad
\label{beta1}
\\
\mbox{and}\ 
&&
\Delta^2\langle \beta_{3}^2(x)\rangle
\simeq \frac{1}{2}\int \frac{{\rm d}^3k}{(2\pi)^3}\frac{T}{\omega_-^2},
\label{beta12}
\eeq
where the fluctuations $\langle \beta_{1,2,3}^2(x)\rangle$ are all logarithmically divergent due to the soft modes in the transverse directions. 
Details of the derivation are given in Appendix~\ref{app:C}. 
Consequently, the low-energy fluctuations wash out the order parameter, i.e., they destroy the off-diagonal long-range order, 
\beq
\langle (\Delta+\delta) U(\beta_i)\phi_0 \rangle =0. 
\eeq

This result implies that a DCDW phase with true long-range order strictly speaking does not exist at non-zero temperature. Such a phase may, however, be realized in a modified form, with a quasi-long-range order (QLRO), analogous to that in the 
Berezinsky-Kosterlitz-Thouless phase in two-dimensional systems~\cite{Berezinsky:1970fr} and in smectic liquid crystals~\cite{de1993physics}. As we discuss in the next section, the quasi-long-range order is characterized by a power-law decay of the order parameter correlation function.

At zero temperature, on the other hand, 
quantum fluctuation are not strong enough to break the modulating order. 
In fact, at $T=0$ the second order fluctuations are given by the infrared convergent integral 
 $\Delta^2\langle \beta_{1,2 \ \!\!(3)}^2(x)\rangle \simeq \frac{1}{4}\int\!\frac{{\rm d}^3k}{(2\pi)^3}\frac{1}{\omega_{k \ \!\!(-)}}$,   
 obtained by taking the zero temperature limit of Eqs.~(\ref{C6}) and (\ref{C12}) in Appendix~\ref{app:C}. 
 
The results of this section imply that the transition temperature of the true DCDW phase is $T_{DCDW}=0$. Now assume that there is a critical temperature $T_{c}>0$, where the system becomes unstable with respect to the formation of a state with a modulated order parameter. Then a quasi-one-dimensionally ordered phase or a phase with true long-range order in two or three dimensions may be realized below $T_{c}$~\cite{Baym:1982ca}. To determine which phase is preferred, one must, in principle, compare their free energies. Considering the different nature of these phases, this is a challenging task.

\subsection{Long-range correlations}
We now explore the behavior of the correlation functions 
in the Gaussian approximation. 
Since the order parameter is vector-like, 
we define correlation functions among the components: 
\beq
f_{ij}(x)=\langle \phi_i(x)\phi^*_j(0)\rangle. \label{lrcf}
\eeq
These correlation functions are spatially anisotropic owing to the one-dimensional modulation of the background. 
We compute the dependence of the iso-scalar correlation function on $z$.
The diagonal components which contribute to the scalar channel are of the form: 
\beq
f_{11}(\hat{z}z)
&\simeq& 
\frac{1}{8}\Delta^2 \cos{qz} e^{-\sum_{i=1,2,3}\langle ({\beta_i^-})^2\rangle/2}, 
\eeq
where $\beta_i^{-}\equiv \beta_i(z) - \beta_i(0)$. 
For details we refer to Appendix~\ref{app:D}. 

Similar results are obtained 
for the other components: 
\beq
f_{22}(\hat{z}z)
&\simeq& 
\frac{1}{8}\Delta^2 \cos{qz} e^{-\sum_{i=1,2,3}\langle ({\beta_i^-})^2\rangle/2}, 
\\
f_{33}(\hat{z}z)
&\simeq& 
\frac{1}{4}\Delta^2 \cos{qz} e^{-\sum_{i=2,3}\langle ({\beta_i^-})^2\rangle/2}, 
\\
\mbox{and} \ \ 
f_{44}(\hat{z}z)
&\simeq& 
\frac{1}{2}\Delta^2 \cos{qz} e^{-\langle ({\beta_3^-})^2\rangle/2}. 
\eeq
Here the exponents, 
$\langle ({\beta_{i}^-})^{2}\rangle$, 
exhibit the following functional form at large $z$, 
\beq
\langle ({\beta_{1,2\ \!\!(3)}^-})^{2}\rangle/2
&\simeq&
\frac{T}{2\Delta^2}\int \frac{{\rm d}^3 k}{(2\pi)^3}
\frac{1 - \cos{k_zz}}{\omega_{k\ \!\!(-)}^2} \nn
&\simeq& 
\frac{T}{16\pi a_{6,1}\Delta^2u_{z-}}\ln{\frac{z\Lambda^2}{2q}}, 
\eeq
where $\Lambda$ is an ultraviolet cutoff. 

Putting it all the together, we obtain the long-range scalar correlation in the $z$ direction, 
\beq
\langle \phi(z\hat{z})\cdot\phi^*(0)\rangle 
&\sim & \frac{1}{2}\Delta^2 \cos{qz}\lk\frac{z}{z_0}\rk^{-T/T_0},
\eeq
where $z_0 \equiv 2q/\Lambda^2$, and $T_0 \equiv 16\pi a_{6,1}\Delta^2u_{z-}$.
In the similar manner, we compute the form of the long-range correlation function in transverse directions,
\beq
\langle \phi(x_{t}\hat{x}_{t})\cdot\phi^*(0)\rangle 
&\sim & \frac{1}{2}\Delta^2 \lk\frac{x_t}{x_0}\rk^{-2T/T_0} , 
\eeq
where $x_0 \equiv \Lambda^{-1}$, 
$x_t$ is the transverse distance,
and the factor $2$ in the exponent of ${x_t}/{x_0}$ reflects the number of transverse directions.
Note that, in contrast to the longitudinal direction, there is no modulation of the correlation function in the transverse directions.

In this section we have shown that quasi-long-range order of the one-dimensional DCDW phase feature algebraically decaying correlation functions at large distances. The slow decay of the spatial correlations distinguish the quasi-ordered phase from  normal or disordered phases, characterized by exponential decays. Depending on the experimental resolution and finite size effects, the algebraic correlations can effectively mimic true long-range order~\cite{Berezinsky:1970fr,AlsNielsen:1980zz,Baym:1982ca}.

\section{Summary and outlook}
\label{sec:5}
In this paper we have explored the soft modes of an inhomogeneous chiral condensed phase with one-dimensional modulation, the DCDW phase.
We found that this phase exhibits a flavor-translation locking symmetry and clarified the counting of Nambu-Goldstone modes.
The dispersion relations for collective excitations, including the NG modes, 
were derived. The low-energy modes  
are spatially anisotropic and particularly soft in the directions transverse to the modulation,
owing to the lack of terms quadratic in the transverse momentum in the dispersion relations.
As in smectic liquid crystals, the absence of such terms is a consequence of the symmetry under rotations about any axis orthogonal to the modulation direction~\cite{landau1969statical,de1993physics}. 

Moreover, we have shown that at non-zero temperatures the DCDW phase exhibits a Landau-Peierls instability, i.e., the long range order  is destroyed by low-energy (long-wavelength) fluctuations of the order parameter. Nevertheless, a phase similar to the smectic phases of liquid crystals, characterized by a quasi-long-range order with algebraically decaying order parameter correlation functions, is possible. Such an ``algebraic order'' can, depending on the conditions, emulate true long-range order. In particular, this would be the case, in a finite systems, where the range of the order-parameter correlations exceeds the size of the system.

The experimental verification of ``algebraic order'' can be challenging. The slow decay of the correlations has been observed by light scattering in smectic-A liquid crystals~\cite{AlsNielsen:1980zz}, by neutron scattering in Bragg glass~\cite{Nattermann:1990} and only recently in a two-dimensional system of the Berezinsky-Kosterlitz-Thouless type~\cite{Nitsche:2014xxx}, by measuring the coherence of photons emitted in quasiparticle decays. 
Whether the quasi-one-dimensionally ordered DCDW phase could be observed by an appropriate choice of probes 
is still an open question.
Hence, it would be important to systematically explore how the collective modes 
in the DCDW phase interact with external probes such as 
hadrons (quarks) and photons (gauge fields).

There are also several theoretical issues, that deserve further study. 
In particular, it is known that inhomogeneous chiral phases are favored 
in systems with vector-vector type interactions,
which tend to enhance the size of the inhomogeneous area~\cite{Carignano:2010ac},
and that in the presence of an external magnetic field 
the FF type phase is stabilized, also at finite temperatures, by topological aspects~\cite{Frolov:2010wn,Tatsumi:2014wka}. 
Moreover, since two- and three-dimensional condensates with true long-range order are allowed at any temperature~\cite{landau1969statical}, it would be important to compare the free energy of such phases with that of a one-dimensional condensate.
It would also be interesting to understand how higher order interactions among the collective modes modify the soft modes.
These may affect the Landau-Peierls instability of inhomogeneous phases discussed here. 

Finally, the topics discussed here may have an impact on the physics of compact stars. It has been speculated that various spatially inhomogeneous phases, like nuclear pasta phases~\citep{Nakazato:2009ed} and hadron-quark mixed phases~\cite{Yasutake:2009sh}, could be realized in  the interior of such stars. These could have phenomenological implications, allowing, e.g., novel cooling scenarios~\cite{Tatsumi:2014cea}.
Thus, it would be interesting to study the properties of inhomogeneous chiral condensed phases under conditions relevant for neutron stars in general and compact stars with quark cores in particular, i.e., in charge neutral matter in $\beta$ equilibrium but also at nonzero isospin density~\cite{Abuki:2013vwa} and finite strangeness~\cite{Moreira:2013ura}. 
These topics will be considered in future works.

TGL would like to thank R.~Yoshiike, T.~Maruyama, K.~Iida, and K.~Kamikado 
for useful comments and discussions.
This work is partially supported by
Grant-in-Aid for Scientific Research on Innovative Areas thorough No.~24105008 provided by MEXT. 



\appendix

\section{Effective action of fluctuations}
\label{app:A}
We derive the effective action of the fluctuations to the second order of field expansion. 
It mostly comes from derivative terms: $(\nabla \phi)^2$ and $(\nabla^2 \phi)^2$. 
First of all, we rewrite the field as 
\beq
\phi(x) = (\Delta+\delta) U(\beta_i) \phi_0(x) 
= (\Delta+\delta)V(\beta_i) \hat{\phi}_0,
\eeq
where 
$V\equiv U S$ with $S=e^{qz L_3}$, 
and $\hat{\phi}_0=(1,0,0,0)^T$.

\subsection{$(\nabla \phi)^2$ term}
We derive the derivative term $(\nabla \phi)^2$ up to the second order of fluctuation fields:
\beq
\nabla \phi \cdot \nabla \phi
&=& 
\hat{\phi}_0^T 
\ldk \nabla \delta V^{-1}+(\Delta + \delta) \nabla V^{-1} \rdk 
\ldk \nabla \delta V+(\Delta + \delta) \nabla V \rdk
\hat{\phi}_0 
\nn
&\simeq& 
(\nabla \delta)^2+q^2\delta^2
+2\Delta \delta \lk q^2+2q\nabla_z\beta_3\rk ,
\eeq
where 
\beq
\hat{\phi}_0^T \ldk \nabla V^{-1} \nabla V \rdk \hat{\phi}_0
&=&
q^2 + 2q\nabla_z\beta_3+q^2\beta_3^2+(\nabla\beta_3)^2 ,
\eeq
and we have used the stationary condition under which $a_{4.2}$, $a_{6.1}$, and $a_{6.2}$ terms disappear.

\subsection{$(\nabla^2 \phi)^2$ term}
We next derive the derivative term $(\nabla^2 \phi)^2$:
\beq
\nabla^2 \phi \cdot \nabla^2 \phi
&=& 
\hat{\phi}_0^T 
\ldk \nabla^2(\delta V^{-1})+\Delta \nabla^2 V^{-1} \rdk
\ldk \nabla^2(\delta V)+\Delta \nabla^2 V \rdk
\hat{\phi}_0 
\nn
&=& 
\langle \nabla^2(\delta V^{-1})\nabla^2(\delta V) \rangle 
+ \Delta \langle \nabla^2(\delta V^{-1}) \nabla^2V \rangle \nn
&& 
+ \Delta \langle \nabla^2V^{-1}\nabla^2(\delta V) \rangle 
+ \Delta^2 \langle \nabla^2 V^{-1}\nabla^2V \rangle,
\eeq
where $\langle \cdots \rangle\equiv\hat{\phi}_0^T \cdots \hat{\phi}_0$.
We expand each term up to second order of the fluctuation fields.
Hereafter we chop off terms which will disappear together with $a_{4.2}$, $a_{6.1}$, and $a_{6.2}$ terms under the stationary condition, and the constants. 
Omitting total derivatives, we obtain 
\beq
\langle \nabla^2(\delta V^{-1})\nabla^2(\delta V)\rangle 
&\simeq&
(\nabla^2\delta-q^2\delta)^2
+4q^2(\nabla_z\delta)^2, \\
\langle \nabla^2(\delta V^{-1}) \nabla^2V \rangle
&\simeq& 
q^4\delta 
+4q \nabla^2\nabla_z\delta \beta_3
-4q^3\nabla_z\delta \beta_3, 
\\
\langle \nabla^2V^{-1}\nabla^2(\delta V) \rangle
&=& 
\langle \nabla^2(\delta V^{-1}) \nabla^2V\rangle, \nn
\langle \nabla^2 V^{-1} \nabla^2 V \rangle
&\simeq& 
\langle \ldk \lk \nabla^2+q^2\rk (S^{-1} \sla{\beta})\rdk 
\ldk \lk \nabla^2+q^2\rk(\sla{\beta}S)\rdk \rangle.
\eeq
where $\sla{\beta}\equiv \sum_{i=1}^3\beta_i L_i$.

\section{Propagators}
\label{app:B}
The following propagators are derived from Eq.~(\ref{disp}): 
\beq
&&
\lk 
\begin{array}{cc} 
\langle \delta(k)\delta^*(k)\rangle & \langle \delta(k)\Delta\beta_3^*(k)\rangle \\ 
\langle \Delta\beta_3(k)\delta^*(k)\rangle & \langle \Delta\beta_3(k)\Delta\beta_3^*(k)\rangle
\end{array}
\rk \nn 
&& \qquad\qquad\qquad
=
\frac{1}{2} \frac{1}{S_{\delta 0}^{-1}(k)S_0^{-1}(k)+g^2(k)}\lk 
\begin{array}{cc}
S_0^{-1}(k) & g(k) \\
-g(k) & S_{\delta 0}^{-1}(k)
\end{array}
\rk , \\
&&
\lk 
\begin{array}{cc} 
\langle \Delta \beta_i(k)\Delta\beta_i^*(k)\rangle 
& \langle \Delta \beta_i(k)\Delta\beta_i^*(k+2q\hat{z})\rangle \\ 
\langle \Delta\beta_i(k+2q\hat{z})\Delta\beta_i^*(k)\rangle 
& \langle \Delta\beta_i(k+2q\hat{z})\Delta\beta_i^*(k+2q\hat{z})\rangle
\end{array}
\rk \nn
&& \qquad\qquad\qquad
=
\frac{1}{2}\frac{1}{S_0^{-1}(k+2q\hat{z})S_0^{-1}(k)-G^2(k)}\lk 
\begin{array}{cc}
S_0^{-1}(k+2q\hat{z}) & -G(k) \\
-G(k) & S_0^{-1}(k)
\end{array}
\rk . 
\eeq
Here note that each component of the latter matrix $S(k)$ corresponds to summing all tree diagrams of the absorption/emission processes, for instance, the $\ltk 1,1\rtk$ component reads 
\beq
S_{11}(k)&=&S_0(k)\sum_{n=0}^{\infty}
\ldk G(k) S_0(k+2q\hat{z}) G(k) S_0(k) \rdk^n \nn
&=&\frac{S_0^{-1}(k+2q\hat{z})}
{S_0^{-1}(k+2q\hat{z})S_0^{-1}(k)-G^2(k)}. 
\eeq

\section{Fluctuation effects on the order parameter}
\label{app:C}
In averaging the order parameter over quadratic fluctuations in Eq.~(\ref{op1}), we have used following results:
\beq
\langle\cos{\lk qz+\beta_3\rk}\rangle
&=&
\cos{qz} \sum_{n=0}^{\infty}\frac{(-1)^n}{(2n)!}\langle \beta_3^{2n} \rangle 
\nn
&=&
\cos{qz} e^{-\langle \beta_3^{2} \rangle/2}, 
\\
\langle\delta \cos{\lk qz+\beta_3\rk}\rangle
&=&
-\sin{qz} \sum_{n=0}^\infty \frac{(-1)^n}{(2n+1)!}
\langle \beta_3^{2n}\rangle (2n+1)\langle \delta \beta_3\rangle 
\nn
&=&
-\sin{qz}\langle \delta \beta_3\rangle e^{-\langle \beta_3^{2} \rangle/2}, 
\eeq
and similarly for the other contributions.

Also, the following finds useful in evaluating integrals like $\int {\rm d}^3{k}~\!T/\omega_\pm^2$,
\beq
\int_{-\infty}^{\infty} {\rm d}k_z
\frac{1}{a k_z^2+b+c k_z^4}
&=& \frac{\pi}{2Ac}\lk \frac{1}{B_-}-\frac{1}{B_+}\rk, 
\eeq
where $(\tilde{a},\tilde{b})=(a, b)/c$, 
$A^2\equiv \frac{\tilde{a}^2}{4}-\tilde{b} >0$, 
and $B_\pm^2\equiv \frac{\tilde{a}}{2}\pm A$. 
In general case where $a=c_1+c_2 k_t^2$, $b=c_3k_t^4$, and $c=1$, for small $k_t$, the above integral results in as follows:
\beq
\frac{\pi}{2Ac}\lk \frac{1}{B_-}-\frac{1}{B_+}\rk 
\simeq \frac{\pi}{c_1}\lk \sqrt{\frac{c_1}{c_3}}\frac{1}{k_t^2} 
-\frac{1}{\sqrt{c_1}} \rk ,
\eeq
where 
\beq
A
=\frac{c_1}{2}+\frac{c_2}{2}k_t^2-\frac{c_3}{c_1} k_t^4+{\mathcal O}(k_t^6), 
%
\mbox{ \ and \ }
B_\pm^2
\simeq 
\ltk\begin{array}{c}
c_1 +c_2k_t^2 \\
\frac{c_3}{c_1} k_t^4
\end{array}\right. . 
\eeq
Here, in a case that $a=4q^2+2k_t^2$, $b=k_t^4$, and $c=1$, which leads to $\int d k_z \omega_{-}^{-2}$, the integral can be evaluated as ${\pi}/{2qk_t^2}$, where $k_t$ is the momentum in the $x$-$y$ plane.

Hereafter we consider the expectation values of second order fluctuations in Eq.~(\ref{op1}).

\subsection{Second order fluctuations for $\delta$ and $\beta_3$}
We evaluate the second order fluctuations by considering infrared (IR) singularities, and in high-temperature and low-energy expansion.

First of all, the second order fluctuations for $\beta_3$ result in as follows:
\beq
S_{3}(k)
&\equiv& 
2\langle \Delta\beta_3(k)\Delta\beta_3^*(k) \rangle
= \frac{S_{\delta 0}^{-1}(k)}{S_{\delta 0}^{-1}(k)S_{0}^{-1}(k)+g^2(k)} 
= \frac{\omega_{+}^2-\omega_{\delta}^2}{\omega_{+}^2-\omega_{-}^2}\frac{1}{\omega^2-\omega_{+}^2}
+ \frac{\omega_{\delta}^2-\omega_{-}^2}{\omega_{+}^2-\omega_{-}^2}\frac{1}{\omega^2-\omega_{-}^2}\nn
&\simeq&
\frac{1}{\omega^2-\omega_{-}^2} \quad \mbox{for small $|\vec{k}|$,}\nn
S_{3}(\tau,\vec{k})
&=&
T \sum_n e^{-i\omega_n \tau} S_{3}(k)
=
\frac{\omega_{+}^2-\omega_{\delta}^2}{\omega_{+}^2-\omega_{-}^2}
\frac{n(\omega_+)e^{\omega_+\tau}+\lk n(\omega_+)+1\rk e^{-\omega_+\tau}}{2\omega_+} \nn
&&+ 
\frac{\omega_{\delta}^2-\omega_{-}^2}{\omega_{+}^2-\omega_{-}^2}
\frac{n(\omega_-)e^{\omega_-\tau}+\lk n(\omega_-)+1\rk e^{-\omega_-\tau}}{2\omega_-}, \label{C6} \\
S_{3}(0,\vec{k})
&\simeq&
\frac{T}{\omega_-^2} \quad \mbox{for small $|\vec{k}|$ and $\omega_\pm/T$,} 
\eeq
where $\omega_{\delta}^2=M^2+a_{6.4}\Delta^2\vec{k}^2+4a_{6.1}q^2(k_z)^2+a_{6.1}(\vec{k}^2)^2$,
and $n(x)=\lk e^{x/T}-1\rk^{-1}$ the Bose distribution function.
Here the $T=0$ limit of (\ref{C6}) reads $S_{3}(0,\vec{k})\simeq\frac{1}{2\omega_-}$.

Secondly, the second order fluctuations for $\delta$ result in as follows:
\beq
S_{\delta}(k)
&\equiv& 
2\langle \delta(k)\delta^*(k) \rangle
= \frac{S_{0}^{-1}(k)}{S_{\delta 0}^{-1}(k)S_{0}^{-1}(k)+g^2(k)} 
= \frac{\omega_{+}^2-\omega_{\beta}^2}{\omega_{+}^2-\omega_{-}^2}\frac{1}{\omega^2-\omega_{+}^2}
+ \frac{\omega_{\beta}^2-\omega_{-}^2}{\omega_{+}^2-\omega_{-}^2}\frac{1}{\omega^2-\omega_{-}^2}\nn
&\simeq&
\frac{1}{\omega^2-\omega_{+}^2} \quad \mbox{for small $|\vec{k}|$,}\nn
S_{\delta}(\tau,\vec{k})
&=&
T \sum_n e^{-i\omega_n \tau} S_{\delta}(k)
=
\frac{\omega_{+}^2-\omega_{\beta}^2}{\omega_{+}^2-\omega_{-}^2}
\frac{n(\omega_+)e^{\omega_+\tau}+\lk n(\omega_+)+1\rk e^{-\omega_+\tau}}{2\omega_+} \nn
&&+
\frac{\omega_{\beta}^2-\omega_{-}^2}{\omega_{+}^2-\omega_{-}^2}
\frac{n(\omega_-)e^{\omega_-\tau}+\lk n(\omega_-)+1\rk e^{-\omega_-\tau}}{2\omega_-}, \label{C8} \\
S_{\delta}(0,\vec{k})
&\simeq&
\frac{T}{\omega_+^2} \quad \mbox{for small $|\vec{k}|$ and $\omega_{\pm}/T$,}
\eeq
where $\omega_{\beta}^2=a_{6,1}\ldk 4q^2 k_z^2 +(\vec{k}^2)^2 \rdk$,
and the $T=0$ limit of (\ref{C8}) reads $S_{3}(0,\vec{k})\simeq\frac{1}{2\omega_+}$.

Finally, the second order fluctuations for $\delta$ and $\beta_3$ result in as follows:
\beq
S_{\delta 3}(k)
&\equiv& 
2\langle \delta(k)\Delta\beta_3^*(k) \rangle
= \frac{g(k)}{S_{\delta 0}^{-1}(k)S_{0}^{-1}(k)+g^2(k)} 
= \frac{g(k)}{\omega_{+}^2-\omega_{-}^2}
\lk \frac{1}{\omega^2-\omega_{+}^2}-\frac{1}{\omega^2-\omega_{-}^2}\rk, \nn
S_{\delta 3}(\tau,\vec{k})
&=&
T \sum_n e^{-i\omega_n \tau} S_{\delta 3}(k)\nn
&=&
\frac{g(k)}{\omega_{+}^2-\omega_{-}^2}
\ldk \frac{n(\omega_-)e^{\omega_-\tau}+\lk n(\omega_-)+1\rk e^{-\omega_-\tau}}{2\omega_-}
- 
\frac{n(\omega_+)e^{\omega_+\tau}+\lk n(\omega_+)+1\rk e^{-\omega_+\tau}}{2\omega_+}
\rdk, \nn \\
S_{\delta 3}(0,\vec{k})
&\simeq&
\frac{g(k)}{\omega_+^2-\omega_-^2}
\ldk \frac{T}{\omega_-^2}-\frac{T}{\omega_+^2}\rdk 
\simeq
\frac{2iq\Delta^2 a_{6.2}k_z}{M^2}\frac{T}{\omega_-^2}
\quad \ \mbox{for small $|\vec{k}|$ and $\omega_{\pm}/T$}.
\eeq

From the above results, the second order fluctuations for $\delta$ and $\beta_3$ in real space read
$\Delta^2 \langle \beta^2_3(x) \rangle = \frac{1}{2}\int\frac{{\rm d}^3k}{(2\pi)^3}S_3(0;\vk)
\simeq \frac{1}{2}\int \frac{{\rm d}^3k}{(2\pi)^3}\frac{T}{\omega_-^2}$,
and $\Delta \langle \delta(x)\beta_3(x) \rangle 
= \frac{1}{2}\int\frac{{\rm d}^3k}{(2\pi)^3}S_{\delta 3}(0;\vk) \simeq 0$.

\subsection{Second order fluctuations for $\beta_1$ and $\beta_2$}
Here for later convenience, we calculate the imaginary time correlations of mean square fluctuation of the $\beta_{1,2}$.

For the $\beta_{1,2}$ with same momentum,
\beq
S_{11}(k)
&\equiv&
2\langle \Delta\beta_{1,2}(k)\Delta\beta_{1,2}^*(k) \rangle 
=\frac{S_{0}^{-1}(k+2q\hat{z})}{S_{0}^{-1}(k+2q\hat{z})S_{0}^{-1}(k)-G^2(k)} 
\nn
S_{11}(\tau,\vec{k})
&=&
T \sum_n e^{-i\omega_n \tau} S_{11}(k)
=\frac{ f_1(\vec{k})}{2\omega_k}\ldk n(\omega_k)e^{\omega_k\tau}
+\lk n(\omega_k)+1\rk e^{-\omega_k\tau}\rdk, \label{C12} \\ 
S_{11}(0,\vec{k})
&\simeq&
\frac{T}{\omega_k^2} \quad \mbox{for small $|\vec{k}|$ and $\omega_k/T$,} \label{C13}
\eeq
where $\omega_k$ is the dispersion relation for $\beta_{1,2}$ given by Eq.~(\ref{disp1}), 
and $f_1(\vec{k})=\lk 1-\frac{Y_{+1}-Y_{-1}}{Y_{+3}-Y_{+1}}\rk^{-2} \simeq 1$ for $|\vec{k}|\simeq 0$ 
with $Y_{\pm n}:=(\vec{k}^2-q^2)^2_{k_z\rightarrow k_z+nq}$. 
Here the $T=0$ limit of (\ref{C12}) reads $S_{11}(0,\vec{k})\simeq\frac{1}{2\omega_k}$.

Similarly, for the $\beta_{1,2}$ with different momentum,
\beq
S_{12}(k)
&\equiv&
2\langle \Delta\beta_{1,2}(k)\Delta\beta_{1,2}^*(k+2q\hat{z}) \rangle 
=\frac{-G(k)}{S_{0}^{-1}(k+2q\hat{z})S_{0}^{-1}(k)-G^2(k)} 
\nn
S_{12}(\tau,\vec{k})
&=&
T \sum_n e^{-i\omega_n \tau} S_{12}(k)
=\frac{ f_2(\vec{k})}{2\omega_k}\ldk n(\omega_k)e^{\omega_k\tau}+\lk n(\omega_k)+1\rk e^{-\omega_k\tau}\rdk, \label{C14} \\
S_{12}(0,\vec{k}) 
&\simeq&
\frac{k_z\vec{k}^2}{8q^3} \frac{T}{\omega_k^2} \quad \mbox{for small $|\vec{k}|$ and $\omega_k/T$,}
\eeq
where 
$f_2(\vec{k})=\frac{Y_{+1}-Y_{-1}}{Y_{+3}-Y_{+1}}
\lk 1-\frac{Y_{+1}-Y_{-1}}{Y_{+3}-Y_{+1}}\rk^{-2} 
\simeq \frac{k_z\vec{k}^2}{8q^3}$ for $|\vec{k}|\simeq 0$.
Here $S_{12}(0,\vec{k})\simeq\frac{k_z\vec{k}^2}{8q^3}\frac{1}{2\omega_k}$ for $T=0$ limit of (\ref{C14}).

From the above results, the real space fluctuation reads
\beq
\Delta^2 \langle {\beta_{i=1,2}}^{\!\!\!\!\!\!\!\!\!\!\!\!\!\!2\ \ \ \ }(x)\rangle
&=&
\sum\!\!\!\!\!\!\!\int\sum\!\!\!\!\!\!\!\int
{\rm d}p {\rm d}k e^{i(p+k)x} 
\langle \Delta \beta_i(p)\Delta \beta_i(k)\rangle\nn
&=&
2 \cos{2qz} 
\sum\!\!\!\!\!\!\!\int {\rm d}k
{\rm Re}\langle \Delta\beta_i^*(k+2q\hat{z})\Delta\beta_i(k)\rangle \nn
&&+2\sin{2qz} 
\sum\!\!\!\!\!\!\!\int {\rm d}k
{\rm Im}\langle \Delta\beta_i^*(k+2q\hat{z})\Delta\beta_i(k)\rangle 
+ 
\sum\!\!\!\!\!\!\!\int {\rm d}k \langle \Delta\beta_i(-k)\Delta\beta_i(k)\rangle \nn
&=&
2 \cos{2qz} \int \frac{{\rm d}^3k}{(2\pi)^3} \frac{S_{21}(0;\vec{k})}{2}
+\int \frac{{\rm d}^3k}{(2\pi)^3} \frac{S_{11}(0;\vec{k})}{2}, 
\eeq
where in the last line the first term gives zero due to the odd function, and the second diverges logarithmically at $k=0$ at finite temperature.

\section{Order parameter correlations}
\label{app:D}
The following integral is useful in evaluation of order parameter correlation functions at long range, 
\beq
\int_{-\infty}^{\infty} {\rm d}k_z
\frac{\cos(k_z z)}{a k_z^2+b+c k_z^4}
&=& \frac{\pi}{2cA}\lk \frac{e^{-B_- z}}{B_-}-\frac{e^{-B_+ z}}{B_+}\rk ,
\eeq
where $A^2\equiv \frac{\tilde{a}^2}{4}-\tilde{b} >0$, $B_\pm^2\equiv \frac{\tilde{a}}{2}\pm A$,
and $(\tilde{a},\tilde{b})=(a, b)/c$.
In general case where $a=c_1+c_2 k_t^2$, $b=c_3k_t^4$, 
for small $k_t$, 
the above integral results in as follows:
\beq
&& \frac{\pi}{2cA}\lk \frac{e^{-B_- z}}{B_-}-\frac{e^{-B_+ z}}{B_+}\rk 
\simeq
\frac{\pi}{c_1}\lk \frac{e^{-\sqrt{\frac{c_3}{c_1}}k_t^2 z}}{\sqrt{\frac{c_3}{c_1}}k_t^2 }
-\frac{e^{-\sqrt{\frac{c_1}{c}} z}}{\sqrt{\frac{c_1}{c}}}\rk,
\eeq
where
\beq
A
=\frac{\tilde{c}_1}{2}+\frac{\tilde{c}_2}{2}k_t^2-\frac{\tilde{c}_3}{\tilde{c}_1} k_t^4
+{\mathcal O}(k_t^6), 
\mbox{ \ and \ }
B_\pm^2 
\simeq 
\ltk\begin{array}{c}
\tilde{c}_1 +\tilde{c}_2k_t^2 \\
\frac{\tilde{c}_3}{\tilde{c}_1} k_t^4
\end{array}\right. .
\eeq
With the above results, we can evaluate the following integral,
\beq
&&
\frac{1}{(2\pi)^2}
\int_{-\infty}^{\infty} {\rm d}k_z \int_{0}^{\Lambda}{\rm d}k_t k_t
\frac{1-\cos(k_z z)}{(c_1+c_2k_t^2)k_z^2+c_3k_t^4+ck_z^4} \nn
&& \simeq 
\int_{0}^{\Lambda}\frac{{\rm d}k_t k_t}{(2\pi)^2}
\frac{\pi}{c_1}
\ldk \lk \frac{1}{\sqrt{\frac{c_3}{c_1}}k_t^2 }
-\frac{1}{\sqrt{\frac{c_1}{c}}}\rk 
-\lk \frac{e^{-\sqrt{\frac{c_3}{c_1}}k_t^2 z}}{\sqrt{\frac{c_3}{c_1}}k_t^2 }
-\frac{e^{-\sqrt{\frac{c_1}{c}} z}}{\sqrt{\frac{c_1}{c}}}\rk 
 \rdk \ \mbox{  for IR region of  } k_t\nn
&& \simeq 
\int_{0}^{\Lambda}\frac{{\rm d}k_t }{4\pi}
\frac{ 1
-e^{-\sqrt{\frac{c_3}{c_1}}k_t^2 z}}{c_1\sqrt{\frac{c_3}{c_1}}k_t}
\ \mbox{  for large  } z 
\nn
&& \simeq 
\frac{1}{8\sqrt{c_1c_3}\pi}\ln\lk \sqrt{\frac{c_3}{c_1}}z\Lambda^2\rk. 
\eeq
where 
$\Lambda$ is the ultraviolet cutoff.

Similarly, the following integral is also useful, 
\beq
\int_{-\infty}^{\infty} {\rm d}k_z
\frac{\sin(k_z z)k_z}{a k_z^2+b+ck_z^4}
&=& \frac{\pi}{2cA}\lk e^{-B_- z}-e^{-B_+ z}\rk . 
\eeq
In general case where $a=c_1+c_2 k_t^2$, $b=c_3k_t^4$, 
for small $k_t$, 
we obtain
\beq
\frac{\pi}{2cA}\lk e^{-B_-z}-e^{-B_+z}\rk 
\simeq 
&& \frac{\pi}{c_1}
\lk e^{-\sqrt{\frac{c_3}{c_1}}k_t^2 z}-e^{-\sqrt{\frac{c_1}{c}}z}\rk . 
\eeq
With this result, we can evaluate the following integral,
\beq
&&
\frac{2\pi}{(2\pi)^3}\int_{-\infty}^{\infty} {\rm d}k_z \int_{0}^{\Lambda}{\rm d}k_t k_t
\frac{\sin(k_z z) k_z}{a k_z^2+b+ck_z^4} \nn
&& \simeq 
\int_{0}^{\Lambda}\frac{{\rm d}k_t k_t}{(2\pi)^2}
\frac{\pi}{c_1}
\lk e^{-\sqrt{\frac{c_3}{c_1}}k_t^2 z}-e^{-\sqrt{\frac{c_1}{c}}z}\rk 
\ \mbox{  for IR region of  } k_t\nn
&& \simeq 
\int_{0}^{\Lambda} \frac{{\rm d}k_t}{4 \pi c_1} k_t
e^{-\sqrt{\frac{c_3}{c_1}}k_t^2 z}
\ \mbox{  for large  } z 
\nn
&& \simeq 
\frac{1}{8\pi\sqrt{c_1c_3} }\frac{1}{z}.
\eeq

\subsection{long-range correlations of diagonal components}

In evaluating the long-range correlation functions of diagonal components~(\ref{lrcf}),
the following expectation values
with $\beta_i^{\pm}\equiv \beta_i(z)\pm \beta_i(0)$
are useful:

\beq
\langle \cos(qz+\beta_{i=1,2,3}^{\pm}) \rangle 
&=&\cos{qz}e^{-\langle {\beta_i^\pm}^2\rangle/2},
\\
\langle \delta(x)\cos(qz+\beta_3^{\pm}) \rangle 
&=&
\mp\sin{qz}\langle \delta(x) \beta_{3}(0)\rangle
e^{-\langle \beta_{3\pm}^{2}\rangle/2}\\
\langle \delta(0)\cos(qz+\beta_3^{\pm}) \rangle 
&=&
-\sin{qz}\langle \delta(0) \beta_{3}(x)\rangle
e^{-\langle \beta_{3\pm}^{2}\rangle/2},\\
\langle \delta(x)\delta(0)\cos(qz+\beta_3^{\pm}) \rangle 
&=&
\cos{qz} e^{-\langle \beta_{3\pm}^2\rangle/2} 
\ldk \langle \delta(x) \delta(0)\rangle 
\mp \langle \delta(x) \beta_{3}(0)\rangle \langle \delta(0) \beta_{3}(x)\rangle \rdk, \nn
\eeq
where 
\beq
\Delta \langle \delta(x)\beta_3(0) \rangle
&=&
\Delta 
\sum\!\!\!\!\!\!\!\int{\rm d}k e^{ikx}\langle \delta(k)\beta_3(-k) \rangle
=\frac{1}{2}\int \frac{{\rm d}^3 k}{(2\pi)^3} e^{i\vec{k}\cdot \vec{x}}S_{\delta 3}(0;\vec{k}) \nn
&\simeq& 
\frac{1}{2}\int \frac{{\rm d}^3 k}{(2\pi)^3} i\sin{(k_zz)}
\frac{2iq\Delta^2 a_{6.2}k_z}{M^2}\frac{T}{\omega_-^2}
\quad \mbox{ for} \ x \parallel \hat{z} \nn
&\simeq& 
-\frac{a_{6.2}q\Delta^2 T}{8\pi a_{6.1}M^2}\frac{1}{qz}, \\
\langle \delta(0)\beta_3(x) \rangle
&=&-\langle \delta(x)\beta_3(0) \rangle, \\
\langle {\beta_{3}^\pm}^2\rangle/2
&=&
\sum\!\!\!\!\!\!\!\int{\rm d}k (1\pm e^{ik_zz}) \langle \beta_3(k) \beta_3^*(k)\rangle 
=
\frac{1}{2\Delta^2}
\int \frac{{\rm d}^3 k}{(2\pi)^3}
\lk 1\pm \cos{k_zz}\rk S_{3}(0;\vec{k}) 
\nn
&\simeq&
\frac{T}{2\Delta^2}\int \frac{{\rm d}^3 k}{(2\pi)^3}
\frac{1\pm \cos{k_zz}}{\omega_-^2}
\nn
&=&
\ltk 
\begin{array}{l}
\infty \\
\sim \frac{T}{16\pi a_{6,1}\Delta^2u_{z-}}\ln{\frac{z\Lambda^2}{2q}}
\end{array}
\right. , 
\eeq
\beq
\langle {\beta_{i=1,2}^\pm}^{\!\!\!\!\!\!\!\!\!2\ \ }\rangle/2
&=&
\sum\!\!\!\!\!\!\!\int{\rm d}k
\ldk \langle \beta_i(k) \beta_i^*(k)\rangle (1\pm e^{ik_zz}) 
\pm \lk \langle \beta_i(k) \beta_i^*(k+2q\hat{z})\rangle e^{ik_zz}+c.c.  \rk 
\rdk \nn
&=&
\frac{1}{2\Delta^2}\int \frac{{\rm d}^3 k}{(2\pi)^3}
\ldk 
\lk 1\pm \cos{k_zz}\rk S_{11}(0;\vec{k}) 
\pm \lk \sin{k_zz}{\rm Im}S_{12}(0;\vec{k}) +\cos{k_zz}{\rm Re}S_{12}(0;\vec{k})\rk 
\rdk \nn
&\simeq&
\frac{T}{2\Delta^2}\int \frac{{\rm d}^3 k}{(2\pi)^3}
\frac{1\pm \cos{k_zz}}{\omega_k^2}\nn
&=&
\ltk 
\begin{array}{l}
\infty \\
\sim \frac{T}{16\pi a_{6,1}\Delta^2u_{z-}}\ln{\frac{z\Lambda^2}{2q}}
\end{array}
\right. . 
\eeq
From the above results, for instance, the $\ltk 1,1\rtk$ component in the $z$ direction reads
\beq
f_{11}(\hat{z}z)
&=& 
\langle (\Delta+\delta(z))(\Delta+\delta(0))
\cos(qz+\beta_3(z))\cos\beta_3(0) \nn
&&
\hspace{1.85in} \times
\cos\beta_2(z)\cos\beta_2(0)
\cos\beta_1(z)\cos\beta_1(0)\rangle 
\nn
&=&
\frac{1}{8}\langle (\Delta+\delta(z))(\Delta+\delta(0))
\ldk \cos(qz+\beta_3^{+})+\cos(qz+\beta_3^{-}) \rdk \rangle 
\nn
&&
\hspace{1.85in} \times
\langle \lk \cos\beta_2^{+}+\cos\beta_2^{-} \rk 
\lk \cos\beta_1^{+}+\cos\beta_1^{-} \rk \rangle 
\nn
&\simeq&
\frac{1}{8}e^{-\sum_i\langle {\beta_i^-}^2\rangle/2}
\ldk 
\Delta^2 \cos{qz}\lk1- \langle \delta(z) \beta_3(0)\rangle^2\rk 
+2\Delta \sin{qz}\langle \delta(z) \beta_3(0)\rangle 
\rdk
\nn
&\simeq& 
\frac{1}{8}\Delta^2 \cos{qz} e^{-\sum_{i=1,2,3}\langle {\beta_i^-}^2\rangle/2},
\eeq
where we have used the fact that $\langle {\beta_i^+}^2\rangle$ are logarithmically divergent, 
and terms including $\langle \delta(z) \beta_3(0)\rangle \propto z^{-1}$ drop faster than others for a large distance in the $z$ direction.
Also, $\langle \delta(z) \delta(0)\rangle$ corresponds to a massive mode, which does not contribute to the long-range correlations.

Similarly, in the $x$-$y$ directions, we can obtain the following results: 
\beq
\langle \cos\beta_{i=1,2,3}^{\pm} \rangle 
&=& e^{-\langle {\beta_i^\pm}^2\rangle/2}, \\
\langle \delta(x_{t})\cos\beta_3^{\pm} \rangle 
&=& 0, \\
\langle \delta(0)\cos\beta_3^{\pm} \rangle 
&=& 0, \\
\langle \delta(x_{t})\delta(0)\cos\beta_3^{\pm} \rangle 
&=&
e^{-\langle \beta_{3\pm}^2\rangle/2} 
\ldk \langle \delta(x_{t}) \delta(0)\rangle 
\mp \langle \delta(x_{t}) \beta_{3}(0)\rangle \langle \delta(0) \beta_{3}(x_{t})\rangle \rdk
\eeq
where $\beta_i^{\pm}\equiv \beta_i(x_t)\pm \beta_i(0)$, and
\beq
\langle {\beta_{i=1,2,3}^\pm}^{\!\!\!\!\!\!\!\!\!\!\!\!\!\!2\ \ \ \ }\rangle/2
&=&
\ltk 
\begin{array}{l}
\infty \\
\sim \frac{T}{16\pi a_{6,1}\Delta^2 u_{z-}}\ln{(x_t \Lambda)^2}
\end{array}
\right. .
\eeq

\end{document}